\pdfpageattr{/Group <</S /Transparency /I true /CS /DeviceRGB>>}

\documentclass{iau}
\usepackage{graphicx}

\title[Spin-orbit angle in compact planetary systems: the 55 Cancri system case] 
{Spin-orbit angle in compact planetary systems perturbed by an inclined
companion. \\ Application to the 55 Cancri system}

\author[Gwena\"el Bou\'e \& Daniel C. Fabrycky]   
{Gwena\"el Bou\'e$^1$ \and Daniel C. Fabrycky$^2$}

\affiliation{$^1$Sorbonne Universit\'es, UPMC Univ Paris 06, UMR 8028,
ASD-IMCCE, \\ Observatoire de Paris, F-75014 Paris, France \\
email: {\tt boue@imcce.fr} \\[\affilskip]
$^2$Department of Astronomy and Astrophysics, University of Chicago, \\
5640 South Ellis Avenue, Chicago, IL 60637, USA \\ email: {\tt
fabrycky@uchicago.edu}}

\pubyear{2014}
\volume{310}  
\pagerange{119--126}
\jname{Complex Planetary Systems}
\editors{A.C. Editor, B.D. Editor \& C.E. Editor, eds.}
\begin{document}

\maketitle

\begin{abstract}
The stellar spin orientation relative to the orbital planes of
multiplanet systems are becoming accessible to observations. For
example, 55 Cancri is a system composed of 5 planets orbiting a member
of a stellar binary for which a projected obliquity of
$ 72\pm12 ^\circ$ relative to the orbit of the innermost
planet has been reported (\cite[Bourrier \& H\'ebrard
2014]{Bourrier14}). This large obliquity has been attributed to the
perturbation induced by the binary. Here we describe the secular
evolution of similar systems and we discuss the case of the 55
Cancri system more deeply. We provide two different
orbital configurations compatible with the currently
available observations.

\end{abstract}

\firstsection 
\section{Introduction}
Consider a gravitational system composed of a central star, several
planets on relatively tight orbits, and a perturbing body at a large
semimajor axis. Any inclination of the outermost orbit relative to the
inner ones would produce a long term precession of the planet orbital
planes.  Moreover, if the planets are close enough to each other, they
all tilt in concert and their secular motion can be described as a
solid rotation of the whole planet system (\cite[Innanen \etal{}
1997]{Innanen97}). 
In the following, this constraint is assumed to be fulfilled.

The goal of the present study is to determine the full range of spin-orbit
angle that can be generated by the secular precession motion described
above. By definition, the spin-orbit angle -- also called the
stellar obliquity or simply the obliquity -- is the angle between the
spin-axis of the central star and the normal of the planet plane. This
specific problem has been analyzed extensively in \cite[Bou{\'e} \& Fabrycky
(2014)]{Boue14}. Here, the method is applied to 55 Cancri, a 5 planet system in
a stellar binary.

This work is motivated by a recent measurement of the projected
spin-orbit angle of 55 Cancri performed by \cite[Bourrier \& H{\'e}brard
(2014)]{Bourrier14} using the Rossiter-McLaughlin effect. Their
observations suggest a highly misaligned system with a projected
obliquity $\lambda={72\pm12}^\circ$. This measurement is, however,
disputed by \cite[Lopez-Morales \etal{} (2014)]{Lopez-Morales14}.

\section{Mathematical description of the problem}
The configuration of the system is uniquely defined by three unit
vectors $\vec s$, $\vec w$, and $\vec n$ along the angular momentum of
the stellar rotation, of the planet orbital motion, and of the
companion's orbit, respectively. Their motion is governed by the
following secular equations (\cite[Bou{\'e} \& Fabrycky 2014]{Boue14})
\begin{eqnarray}
\frac{d\vec s}{dt} &=& -\nu_1 (\vec s\cdot \vec w) \vec w \times \vec s, \nonumber \\[0.5em]
\frac{d\vec w}{dt} &=& -\nu_2 (\vec s\cdot \vec w) \vec s \times \vec w 
\label{eq.motion}
                       -\nu_3 (\vec n\cdot \vec w) \vec n \times \vec w, \\[0.5em]
\frac{d\vec n}{dt} &=& -\nu_4 (\vec n\cdot \vec w) \vec w \times \vec n.  \nonumber
\end{eqnarray}
Thus, the evolution is fully characterized by four secular frequencies
$\nu_1$, $\nu_2$, $\nu_3$, and $\nu_4$, only. These frequencies
represent the speed at which each vector precesses around the others.
The interaction between $\vec s$ and $\vec w$ is due to the oblateness of
the central star induced by its rotation. A similar interaction should
exist between $\vec s$ and $\vec n$, but given its weakness, it is
neglected.

\begin{figure}[h]
\begin{center}
 \includegraphics[width=3.4in]{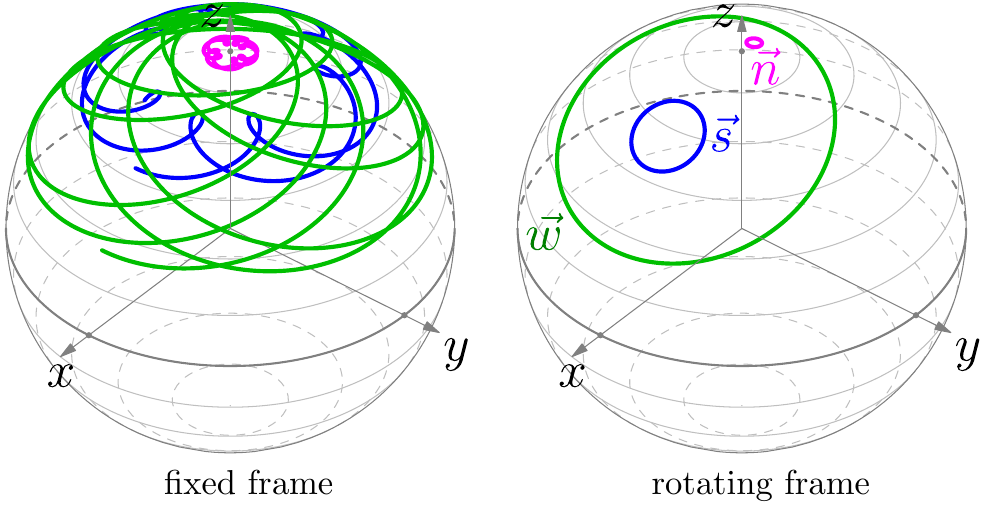} 
 \caption{Typical solution of the equations (\ref{eq.motion}) seen in a
fixed reference frame (left) and in a rotating frame (right). The total
angular momentum is aligned with the $z$-axis.}
   \label{fig1}
\end{center}
\end{figure}

The system described by the differential equations (\ref{eq.motion})
belongs to a class of integrable three vector problems (\cite[Bou{\'e}
\& Laskar 2006]{Boue06}, \cite[2009]{Boue09}). The general solution is a
uniform rotation of the three vectors around the total angular momentum
combined with periodic loops in the rotating frame (see figure \ref{fig1}).

\section{Amplitude of spin-orbit oscillations}
\begin{figure}[t]
\begin{center}
 \includegraphics[width=\linewidth]{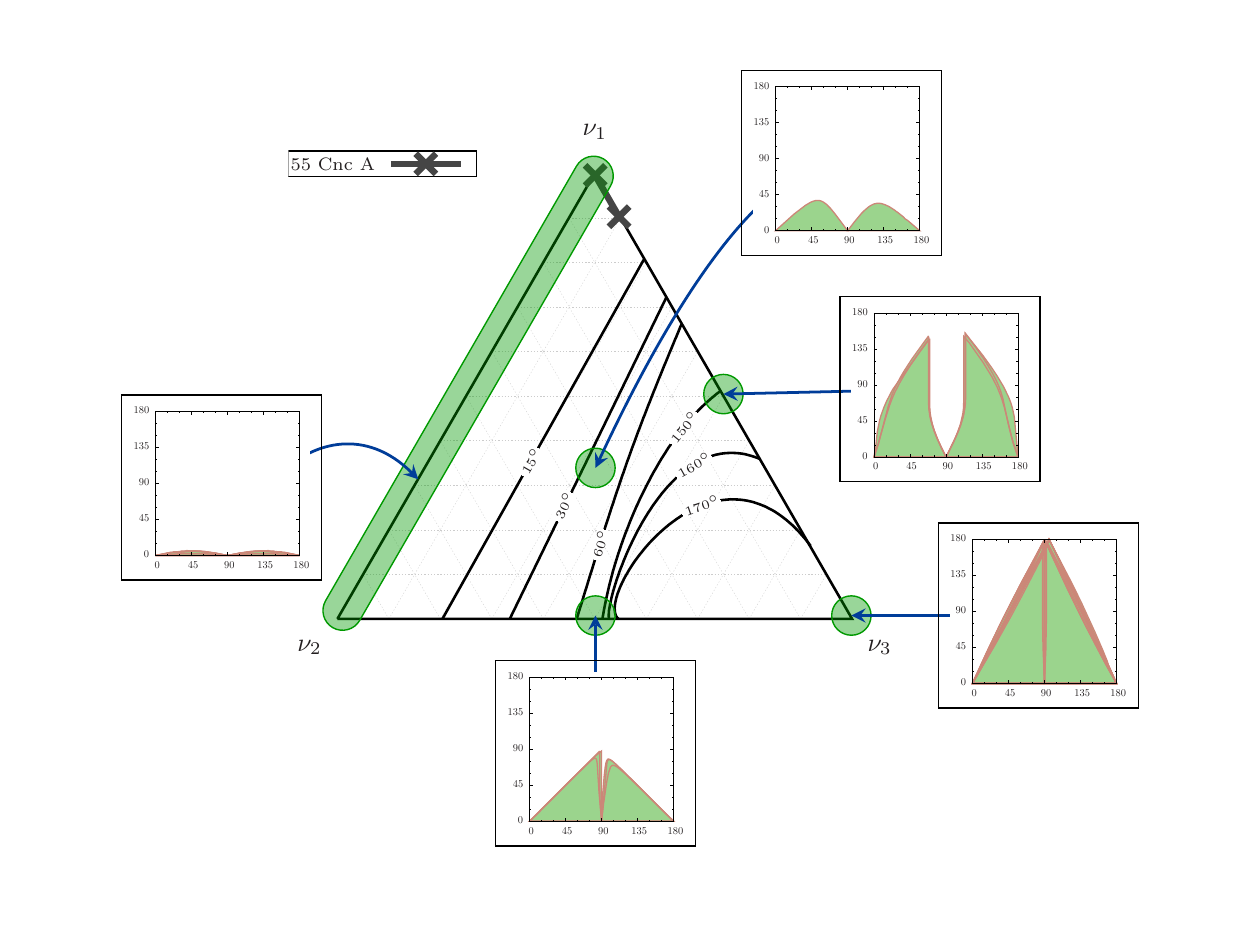} 
 \caption{Ternary diagram with coordinates $(\nu_1, \nu_2, \nu_3)$. The
system 55 Cancri is represented by a segment corresponding to the binary
eccentricity ranging from 0 (top left end) to 0.95 (bottom right end).
Any position in this diagram uniquely determines the behavior of the
stellar obliquity. This value increases from a few degrees at the
$\nu_1$-$\nu_2$ edge up to twice the initial inclination of the binary's
orbit at the $\nu_3$ vertex of the triangle.
Five insets detail the obliquity's full amplitude (vertical axis, in
degrees) as a function of the binary's initial inclination (hortizontal
axis, in degrees), the maximum of which is shown as level curves across
the diagram.
}
   \label{fig2}
\end{center}
\end{figure}

In the specific case where the perturber is a star, the binary's orbit
possesses most of the total angular momentum. By consequence, its plane
is practically invariant and the precession frequency $\nu_4$ is
negligible with respect to the other frequencies. 
Furthermore, if we discard the timescale of the evolution, the three
remaining frequencies can be normalized by an arbitrary constant so that
the effective dimension of the parameter space becomes equal to two.
Using barycentric coordinates $(\nu_1/\nu_\mathrm{tot},
\nu_2/\nu_\mathrm{tot}, \nu_3/\nu_\mathrm{tot})$, where
$\nu_\mathrm{tot} = \nu_1+\nu_2+\nu_3$, all systems with identical
spin-orbit behavior but different timescales can be represented by a
single point in a ternary diagram (Figures \ref{fig2} and 3). 

In this diagram, close to the upper corner, as in the 55 Cancri case, the
frequency $\nu_1$ is the highest, thus the evolution is dominated by the
precession of the central star around the planet plane. Near the bottom
left vertex, the strongest motion is the precession of the planets with
respect to the equator of the star at the frequency $\nu_2$. Finally, in
the vicinity of the bottom right corner where $\nu_3$ dominates, the
evolution is mainly characterized by the precession of the planets
around the binary's orbital plane.

As a result, all systems in the neighborhood of the $\nu_1$-$\nu_2$
edge are lead by a strong coupling between the central star and the planets.
Hence, if the spin-orbit angle is initially small, it remains small.
Conversely, close to the $\nu_3$ vertex the precession of the planets
around the binary's orbit is too fast for the star to follow, thus the
obliquity increases up to twice the initial inclination of the binary.
The middle of the $\nu_1$-$\nu_3$ edge is particular because it
corresponds to a secular spin-orbit resonance in the vicinity of a
Cassini state. In this case, the maximal obliquity can exceed twice the
initial inclination of the binary.

\section{Discussion}
\widowpenalty=500
Using the default values of \cite[Bou{\'e} \& Fabrycky (2014)]{Boue14},
the 55 Cancri system ends up in a region of the parameters space where
no significant spin-orbit misalignment is expected (Figure~\ref{fig2}).
Yet, these parameters are the same as those considered by \cite[Kaib
\etal{} (2011)]{Kaib11} who inferred a most probable projected obliquity
of about 60$^\circ$, in very good agreement with \cite[Bourrier \&
H{\'e}brard (2014)]{Bourrier14} observations. The apparent discrepancy
among the two dynamical analyses is due to the spin-orbit coupling
between the star and the planets which was not taken into account in
\cite[Kaib \etal{} (2011)]{Kaib11}. By consequence, $\nu_1$ and $\nu_2$
were forced to zero and the system was artificially shifted at the
$\nu_3$ corner in Figure \ref{fig2}.

In order to match the Rossiter-McLaughlin observations, it is necessary to
enhance the coupling between the planets and the companion. One solution
is to increase the eccentricity of the binary above $e' \gtrsim 0.987$
(\cite[Bou{\'e} \& Fabrycky 2014]{Boue14}). With this value, the system
reaches the Cassini regime (middle of the $\nu_1$-$\nu_3$ edge in Figure
\ref{fig2}) and the obliquity can grow by a large amount. 
\widowpenalty=150

Another solution is to assume that planet d is the
actual perturber of the inner four planets.
Indeed, planet d
is well separated from the inner ones; 
its inclination with respect to the plane of the sky is estimated
to be $i_d=53\pm 7^\circ$ (\cite[McArthur \etal{} 2004]{McArthur04})
while planet e's inclination is close to
$i_e=90^\circ$ (\cite[Winn \etal{} 2011]{Winn11}), implying that these two planets are mutually inclined\footnote{\cite[Nelson
\etal. (2014)]{Nelson14} showed if the middle three planets are in the
plane of the outermost, then Kozai cycles could destroy the innnermost
planet; however, we suggest the middle three lie near the plane of the
innermost.}; 
          moreover, with this
hypothesis the system falls in the pure orbital regime $(\nu_3 \sim \nu_4) \gg (\nu_1, \nu_2)$ compatible with large spin-orbit angles (\cite[Bou{\'e} \& Fabrycky 2014]{Boue14}).
Nevertheless, this scenario does not provide any hint to explain the
origin of the required large inclination of planet d.

\section{Conclusion}
55 Cancri is a multiplanet system whose central star possesses a
binary companion. This particular hierarchy can produce a misalignment of the
stellar spin-axis with respect to the normal of the planets' plane. The
secular evolution of such systems is uniquely determined by their position
in a ternary diagram in which coordinates are three precession
frequencies $(\nu_1,\nu_2,\nu_3)$. The 55 Cancri system is located in a
region of the diagram with no excitation of the stellar obliquity in
contradiction with recent Rossiter-McLaughlin measurements.
Nevertheless, two different orbital configurations in agreement with the
currently available observations are able to solve this issue. In the
first one, the eccentricity of the stellar companion is set to a high
value: $e'\gtrsim 0.987$. In the second, the outermost planet takes the
place of the perturber.

\appendix

\section{Additional figure}
\begin{figure}[h]
\begin{center}
 \includegraphics[width=\linewidth]{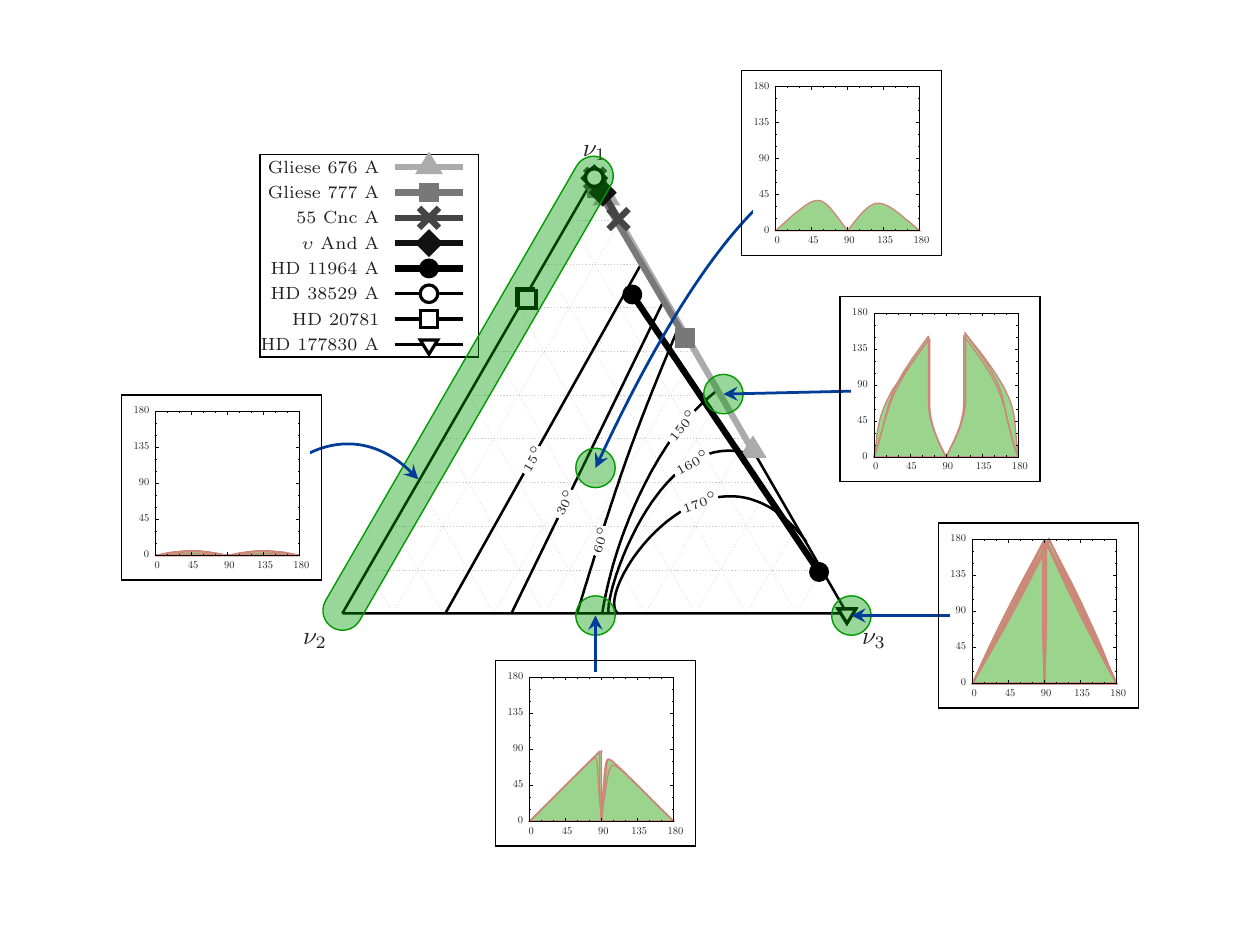} 
 \caption{Same as Figure~\ref{fig2} but with all compact S-type
multiplanet systems in binairies taken from the Open Exoplanet Catalogue
(\cite[Rein 2012]{Rein12}).}
   \label{fig22}
\end{center}
\end{figure}

\end{document}